\newcommand{\ARAA}{ARA\&A}
\newcommand{\AJ}{AJ}
\newcommand{\AaA}{A\&A}
\newcommand{\ApJ}{ApJ}
\newcommand{\MNRAS}{MNRAS}
\newcommand{\Natur}{Nature}

\documentstyle[epsf]{aa}
\begin{document}
\def\gtsima{$\; \buildrel > \over \sim \;$}
\thesaurus{}
\title{Sunyaev-Zel'dovich constraints from black hole-seeded proto-galaxies}
\author{Nabila Aghanim\inst{1} \and Christophe Balland\inst{1} \and Joseph Silk\inst{2}}
\offprints{N. Aghanim, aghanim@ias.fr}
\institute{IAS-CNRS, Universit\'e Paris Sud, B\^atiment 121, F-91405 Orsay 
Cedex \and Department of Astrophysics, University of Oxford, NAPL, Keble Rd, Oxford OX13RH}
\date{Received date / accepted date}
\maketitle
\markboth{Sunyaev-Zel'dovich constraints from black hole-seeded 
proto-galaxies}{}
\begin{abstract}
Recent studies of galactic nuclei suggest that most galaxies are 
seeded by super-massive black holes which power the central nucleus. In this 
picture, the proto-galactic object is likely to have undergone a very active 
phase
during which the surrounding medium was shocked and heated up to very high 
temperatures.
We predict the cosmological effects of this scenario in terms of 
the thermal and kinetic Sunyaev-Zel'dovich distortions 
induced on galactic scales by a population of proto-galaxies. These
predictions are compared to the observational limit on the mean
Compton distortion set by the COBE-FIRAS instrument. This enables us to
derive tight constraints on the fraction of proto-galaxies 
seeded by super-massive black holes as well as on the black hole-to-spheroid 
mass ratio.
Finally, we estimate the contribution of such a population to the angular 
power spectrum of the Cosmic Microwave Background temperature anisotropies
on very small angular scales ($l\simeq 10^4-10^5$).

\end{abstract}
\section{Introduction}
Most theories of hierarchical structure formation
are based on the study of the evolution of density perturbations 
under their own gravity. A density fluctuation, which represents an 
over- (or under-) density with respect to the mean matter 
distribution,  
contains both baryonic and dark matter (DM). The baryonic component sinks
into the gravitational potential of the DM halo. It collapses and
cools, resulting in  star formation. In these scenarios, after the 
gravitational collapse of the DM halo, stars are assumed to be
the first objects to form. A structure will thus end up as an 
emitting object after virialisation has occurred.

An alternative picture involves the formation
of a super-massive black hole (BH) that powers the central regions 
of galaxies \cite[]{lynden-bell69}.
Numerous studies have been performed that relate the quasar luminosity function
to galaxy formation scenarios by assuming that the formation of quasars 
(i.e., BH) in the potential well of the DM halos constitutes
one of the phases in the galaxy formation process 
\cite[]{efstathiou88,haehnelt93,nusser93,haiman97}.
Recent observations even suggest that a super-massive
BH may be present in the centres of all galaxies with spheroidal components 
\cite[]{kormendy95}.\par
Several authors have looked at several consequences of the presence of 
massive 
BHs on galaxy formation and evolution \cite[]{haiman97,natarajan98,silk98}.
In this paper, we investigate the cosmological implications of such an 
alternative scenario for the
Cosmic Microwave Background (CMB) anisotropies and spectral distortions.
More specifically, we
study the effects of the outflows, driven by the BH activity, 
on the gas within the seeded proto-galaxy. In fact, the
outflow expands and shock-heats the ambient medium (proto-galactic gas), 
and then interacts with
the inter-galactic medium (IGM). Three regimes of interest may be 
considered: 1) the high density region of the proto-galaxy, 2) the low density 
IGM and 3) the thin compressed layer (four times denser than the IGM)
induced by the front shock. The second and third regimes give results very
similar to those computed in \cite{aghanim96}. 
We thus focus on the first regime, i.e. the localised effects of the BH-driven shock on the gas within the seeded proto-galaxy. This shock-heated
gas will Compton scatter the CMB photons and induce spectral distortions and
temperature anisotropies through the so-called Sunyaev-Zel'dovich effect
\cite[]{suniaev80}. 
The thermal SZ effect depresses the CMB brightness in the 
Rayleigh-Jeans region and increases it above a frequency of about 219 GHz. Its
amplitude represents the integral along the 
line of sight of the electron pressure. It is proportional to the electron 
density $n_e$ and is characterised by the Compton parameter $y$ defined by:
\begin{equation}
y=2\int_0^R\frac{kT}{m_ec^2}\sigma_Tn_e(l)\,dl,
\label{eq:y}
\end{equation} 
where $T$ is the temperature of the gas, $R$ the physical size of the 
structure, $m_e$ the electron mass, $k$ the Boltzmann constant, $c$ the speed 
of light and $\sigma_T$ the Thomson cross section. An additional secondary 
anisotropy arises due to the first-order Doppler effect of the
CMB photons when they scatter on a structure moving with respect to the Hubble 
flow, with radial peculiar velocity $v_r$. This interaction is called the SZ
kinetic effect. It generates an anisotropy, with no specific spectral 
signature, whose amplitude is given by:
\begin{equation}
\frac{\delta T}{T}=\frac{v_r}{c}\times\left( 2\int_0^R\sigma_Tn_e(l)\,dl
\right).
\label{eq:dt}
\end{equation}

Previous work on galaxy formation has evaluated the global distortion of
a population of galaxies in the virialised regime. The global 
Compton parameter was found to be much smaller than the constraints set by the 
FIRAS instrument (Far InfraRed Absolute Spectro-photometer) on board COBE 
(COsmic Background Explorer) \cite[]{fixsen96} 
on the
global SZ distortion of the universe. By 
contrast, our model focuses on a regime in which  
proto-galaxies undergo a BH formation phase that induces larger distortions.
The paper is organised as follows: in \S 2, we model the shock in an individual
structure and give its physical 
characteristics (size and temperature). In \S 3, we compute the predicted
number density of primordial galaxies, using the 
\cite{press74} mass 
function.  In \S 4, we generalise the description of the shock to the whole
population of proto-galaxies and we simulate maps of the induced secondary 
anisotropies. We estimate this contribution to the CMB anisotropies. We also 
compare our predicted global $y$ parameter to the COBE-FIRAS 
value and derive constraints on the model. Conclusions are given in the last 
section.
\section{Modelling the shock}
In the galaxy formation canonical scenario, a galaxy forms in the 
gravitational potential well of a DM halo of mass $M_{halo}$.
Following \cite{silk98},
we consider that each forming galaxy, which is fully described by the
mass $M_{halo}$, hosts a super-massive BH. The fraction of seeded 
proto-galaxies is thus 100\%. For the sake of 
simplicity, we assume that all proto-galaxies are spheroids. A more 
refined and accurate description would take into account morphological 
segregation \cite[]{balland98}.
However, these effects do not introduce
significant differences. We assume that the BH
radiates during its lifetime $t_{BH}$ a fraction $\epsilon_E=0.1\,-\,0.2$ 
\cite[]{natarajan99} 
of its Eddington luminosity ($L_{BH}=\epsilon_E
\,L_{edd}$). The latter is fixed by the BH mass, $M_{BH}$:
\begin{equation}
L_{edd}=\frac{4\pi M_{BH}m_pGc}{\sigma_T},
\end{equation}
where $m_p$ is the proton mass and $G$ the gravitational constant. 
The mass of the BH should be
directly related to the mass of the proto-galaxy it seeds, and thus 
to the fraction of baryonic matter locked up in the spheroid
in which the central BH collapses. We assume a simple
relation of proportionality ($M_{BH}=\epsilon_{BH}\,M_{sph}$) between 
the mass of the BH and the mass of the spheroid $M_{sph}$. \cite{haehnelt93} 
give $\epsilon_{BH}\simeq 10^{-3}$ and they assume that $\epsilon_{BH}$
declines with mass. However, observations
of galactic nuclei \cite[]{kormendy97,magorrian98} 
indicate that
$\epsilon_{BH}$ is relatively constant. Here, we take the value 
$\epsilon_{BH}=2.10^{-3}$ as advocated by \cite{magorrian98} 
and \cite{silk98}. We 
assume that the medium is instantaneously ionised, which is very likely due to
the intense radiation emitted by the central 
BH \cite[]{voit96}. 
We therefore study the propagation 
of a strong shock driven by a mechanical energy which represents some fraction
of the BH luminosity. The first analytic solutions of spherically symmetric 
explosions were given by \cite{taylor50} 
and \cite{sedov59}
and applied to several astrophysical problems
such as stellar winds and supernovae explosions 
\cite[]{ikeuchi83,bertschinger86,koo92,koo92a,voit96}.
Following \cite{voit96}, 
we study the expansion of the 
shock and its effects on the gas within the seeded proto-galaxy. 
We characterise the shock in an expanding universe 
by a set of physical properties (its radius, 
velocity and temperature). Adapting Voit's solutions to our study, the shock 
at a redshift $z<z_*$, where $z_*$ is the redshift at which the BH switches 
on, has a radius $R_s$ which reads as:
\[
R_s\simeq\left(\frac{\pi G E_0}{H_0^4\delta\Omega_0^3}\right)^{1/5} 
\frac{(1+\Omega_0z_*)^{1/5}}{(1+z_*)^{2/5}}\,\times\]
\begin{equation}
\left[1-\left(\frac{1+\Omega_0z}
{1+\Omega_0z_*}\right)^{1/2}\right]^{2/5}(1+z)^{-1}.
\end{equation}
It propagates at a velocity 
\[
v_s=\frac{2}{5}\left[\frac{\pi E_0H_0G}{3\delta\Omega_b}\frac{\Omega_0^2
(1+z_*)^3}{(1+\Omega_0z_*)^{3/2}}\right]^{1/5}\,\times\]
\begin{equation}
\left[1-\left(\frac{1+\Omega_0z}
{1+\Omega_0z_*}\right)^{1/2}\right]^{-3/5}\left(\frac{1+z}{1+z_*}\right).
\end{equation}
In the two previous equations, $E_0=\epsilon L_{BH}t_{BH}$ represents the
mechanical energy which drives the shock. Very little is known about the 
fraction $\epsilon$ which can be on the order of 0.5 or even higher 
\cite[]{natarajan99}.
For the lifetime of the BH in the bright phase, we choose the recent value 
derived by \cite{haiman97}: 
$t_{BH}=10^{6}$ years (our results, however, are not very
sensitive to the exact value of $t_{BH}$).
$\delta$ is the mean over-density of the proto-galaxy (assumed 
to be spherical), $H_0$ is the Hubble constant, $\Omega_0$ is the 
density parameter and $\Omega_b$ is the baryon density in the universe. Zero
subscripts denote present day values. In the following and throughout the 
paper, we use $h=H_0/(100 \mbox{km/Mpc/s})=0.5$ and $\Omega_b=0.06$ 
\cite[]{walker91} and give the results as a function of $\Omega_0$, for
$\Omega_0=1$ and $\Omega_0=0.3$. 
\par
Following \cite{natarajan99}, 
we assume that the BH radiates 10\% of its Eddington luminosity,
i.e. $\epsilon_E=0.1$ and that half of the corresponding energy is mechanical.
Furthermore, we consider that galaxies form at sufficiently high redshifts
($>5$) so that only hydrogen and helium are present in substantial amounts.
We compute both the size and velocity of the shock and find
that the host proto-galaxy is always embedded in the shocked region.
\subsection{Cooling mechanisms}
The temperature of the shocked medium can be directly derived from the 
equipartition of energy: 
\begin{equation}
T\simeq v_s^2\mu m_p/k.
\end{equation} 
Here, $\mu=0.6$ is the mean molecular weight of a
plasma with primordial abundances. The galactic matter can be shock-heated 
up to very high temperatures.
For the most massive galaxies, we find that the matter is heated up to a few 
$10^8$ K, a value comparable to the temperature of the intracluster medium
in galaxy clusters. For these temperatures and redshifts ($>5$), the main 
cooling process at play is bremsstrahlung. Therefore,
the shocked gas looses heat with a 
cooling rate given by a temperature-dependent cooling function $\Lambda(T)$.
We have compared three different cooling functions 
\cite[]{bertschinger86,koo92,voit96} 
given in the literature for temperatures between a few $10^5$ and $10^8$ K. 
We find that the condition required for efficient cooling is always 
satisfied in our redshift range. This means that the cooling time, 
given by $t_{cool}= 5kT/n_H\Lambda(T)$ with $n_H$
the hydrogen number density, is always smaller than the age of Universe.
Cooling is thus very efficient in our picture.
The comparison between the three cooling functions shows that
they all give essentially the same final temperature after cooling. 
We follow \cite{bertschinger86} 
and we find that the 
shocked matter cools down to $T\simeq 10^5$ K at $z\simeq 5$. 
\section{Number counts of primordial galaxies}
In order to quantify the global effect of the formation of 
primordial galaxies on the CMB, we apply our formalism to a synthetic 
population of galaxies with masses $10^9M_{\sun}\leq M\leq10^{12}M_{\sun}$.

We first assume that the galaxy number density traces, within a linear 
bias, the abundance of collapsed DM halos, as 
predicted by the Press--Schechter (PS) mass function \cite[]{press74}.
We use an initial power-law spectrum with an effective spectral index $n=-2$ 
on galaxy scales. We express the amplitude
of primordial matter fluctuations in terms of the {\it rms}
variance in spheres of 8$h^{-1}$ Mpc, $\sigma_8=0.6$ (as cluster-normalised, 
e.g. \cite{viana96}, 
which corresponds to a bias factor $b\approx 1.67$. For $\Omega = 1$, the
set of parameters corresponds to the ``standard'' biased cold DM model, which
does fit neither small- and large-scale velocities \cite{vittorio86} nor
COBE normalisation. However, we take it as a study case for the computations, 
our second model is the low density cosmological model with $\Omega = 0.3$.
In our picture (no stars are formed yet), the spheroid is gaseous. Its
mass is related to the mass of the DM 
halo via $M_{sph}=\frac{1}{3}M_{halo}$. The mass and luminosity of 
the central BH and thus the predicted SZ distortions are 
therefore inferred from $M_{sph}$ ($M_{BH}=\epsilon_{BH}M_{sph}$,
and $\epsilon_{BH}=2.\,10^{-3}$; cf \S 2). 

To compute the kinetic SZ term of a population 
of proto-galaxies, we need an estimate
of their peculiar velocities with respect to the reference frame. 
As suggested by numerical 
simulations \cite[]{bahcall94,moscardini96}, 
we assume that velocities follow a Gaussian distribution. The peculiar 
velocity of each proto-galaxy is drawn from a Gaussian which is completely 
defined by its {\it rms} value $\sigma_v$. In the range of 
redshifts we have adopted, the structures are in the linear regime, so that
$\sigma_v(z)=\sigma_0\,f(z)$, where the redshift dependence of 
the velocities is given by $f(z) $\cite[]{peebles80,peebles93} as a function 
of the cosmological parameters.
In this equation,
$\sigma_0$ is the present-day {\it rms} peculiar velocity. It is related to
the mass variance on mass scale $M$, 
$\sigma(M)=(1.19\Omega_0)^{(n+3)/6}\,\sigma_8M^{-(n+3)/6}$ 
{\bf \cite{mathiesen98}}, where $n$ is the 
index of the power spectrum. The {\it rms}
velocity can thus be computed for each mass scale.
\section{Results and discussion}
The shock-heated gas within proto-galaxies interacts with the CMB photons 
through
the SZ effect (thermal and kinetic). These interactions generate secondary
temperature anisotropies and spectral distortions. We simulate maps of 
the secondary anisotropies generated by a population of seeded
proto-galaxies formed between redshift 5 and 10. The maps have a resolution of 
about $0.2$ arcseconds to resolve the galaxies and contain
$600\times600$ pixels. The number of sources of mass $M$ at redshift $z$ is 
derived from the PS mass function. Their positions are drawn at random
in the map. The $y$ and $\delta T/T$ profiles for, respectively, 
the thermal and the kinetic effects are directly derived from the integration 
of the gas profile $n_e(R)$ along the line of sight (Eqs. \ref{eq:y} and 
\ref{eq:dt}) assuming spherical symmetry. \par
Similarly to the case of galaxy clusters, we assume that in the early stages 
of formation the gas settles into a hydrostatic equilibrium within the DM
potential. A universal density profile is motivated by \cite{navarro96a}.
However, the gas profile may be softer than that of the DM, and moreover
the existence of a central cusp is ``unobserved''  
\cite[]{kravtsov98}.We thus conservatively adopt the
following parametrised profile for the gas distribution:
\begin{equation}
n_e(R)=n_0\left[1+\left(\frac{R}{R_c}\right)^2\right]^{-\alpha},
\label{eq:prof}
\end{equation}
where $n_0$ is the central density. $\alpha$ is left as a free parameter 
describing the steepness of the profile, whereas $R_c$ is identified with a
core radius as in galaxy clusters. On cluster scales, $R_c$ is typically 10 to 
30 times smaller than the cluster virial radius $R_{vir}$. In our model we 
introduce the parameter $p=R_{vir}/R_c$ which we
vary, similarly to clusters, between 10 and 30. The central
density $n_0$ can be derived from the gas mass of the proto-galaxy using the
following equation:
\begin{equation}
M_{sph}\left(\frac{\Omega_b}{\Omega_0}\right)=m_p\mu\int_0^{R_{vir}}n_e(R)
\,4\pi R^2\,dR, 
\end{equation}
where the virial radius of the structure is given by: 
\begin{equation}
R_{vir}=\frac{(G\,M)^{1/3}}{(3\pi\,H_0)^{2/3}}\,\frac{1}{1+z_{coll}},
\end{equation}
for a critical universe. It is
fixed solely by the mass and the collapse redshift $z_{coll}$.\par
We will give the results for a flat model
with no cosmological constant ($\Omega_0=1$) and an open model 
($\Omega_0=0.3$). Varying the cosmological parameters will vary the number of 
proto-galaxies along the line of sight as well as their peculiar velocities. It
will also modify their physical properties, i.e. the size and velocity of the 
shock and thus the gas temperature. The two cosmological models represent the upper and
lower bounds between which all other cosmological models involving a non-zero cosmological constant fall.

\subsection{Compton distortion}
The CMB photons, scattering off the electrons of the ionised hot gas, induce a
spectral distortion whose amplitude is given by Eq. \ref{eq:y}. The FIRAS
experiment  has measured the mean Compton parameter resulting
from all the interactions undergone by the photons. The result is 
$\overline y_{FIRAS}=1.5\,10^{-5}$ \cite[]{fixsen96}. 
This stringent observational limit 
incorporates the (negligible) contribution of the rather cold intergalactic 
medium and that of all other extragalactic signals. Among these signals, there 
is the contribution of the hot ionised gas in galaxy clusters. The global 
distortion induced by intra-cluster gas has been computed 
\cite[]{de_luca95,barbosa96}, 
and  found to be
of the order of a few $10^{-6}$. In addition to galaxy clusters, one has to 
take into account the
contribution of the proto-galaxy population in terms of the overall Compton 
distortion, $\overline y_{PG}$, induced by the scattering of CMB photons on 
the shock-heated gas.
\par\medskip
Based on simulated maps, we predict $\overline y_{PG}$ and we
compare it to the limit set by COBE-FIRAS. Among all the parameters of the
model, there are four major 
quantities that substantially affect the predictions of the mean Compton 
parameter. Two of them, $\alpha$ and $p$, are related to the gas distribution 
(Eq. \ref{eq:prof}).
The two others are the fraction $f$ of BH-seeded proto-galaxies and the 
BH-to-spheroid mass ratio $\epsilon_{BH}$. We compare
our predicted overall distortion to the COBE-FIRAS limit and look
for the combinations of parameters for which our predictions fit the 
observations. This allows us to constrain the assumptions of our model.  \par
For $\alpha=1/2$, $f=100$\%, $\epsilon_{BH}=2.\,10^{-3}$ and 
both cosmological models, we find $\overline y_{PG} \simeq 10^{-4}$
which exceeds the observational value. In order for our prediction to be
reconciled with the COBE-FIRAS limit, $f$ must be only a few percent. This
constraint on $f$ strongly violates the actual observations 
\cite[]{magorrian98,richstone98}. 
$\alpha=1/2$ is thus excluded by the limit on the 
global distortion whatever value we choose for 
$p=R_{vir}/R_c$. \par
For $\alpha=1$ (i.e. an isothermal profile), an $\Omega_0=1$ model, 
$f=100$\% and 
$\epsilon_{BH}=2.\,10^{-3}$, we find $\overline y_{PG}>\overline y_{FIRAS}$ 
whatever we adopt for $p$.
The fraction $f$ must be smaller than 75\% for the prediction to be compatible 
with the observational limit. Again, this fraction is significantly smaller 
than the 95\% advocated by \cite{magorrian98}. 
Such a constraint
could rule out the isothermal profile. However, up to now, $\epsilon_{BH}$ was 
assumed to be constant and equal to $2.\,10^{-3}$. If we now use the lower 
limit of \cite{magorrian98}, 
that is $\epsilon_{BH}=10^{-3}$, 
together with $f=95$\% or higher there is only a marginal agreement for 
$p\sim 30$ between the predicted and measured distortions. In the open 
model case ($\Omega_0=0.3$), we find approximately the same results. 
For $\overline y_{PG}$ to be compatible with $\overline y_{FIRAS}$ if $p=10$, 
the fraction of BH seeded proto-galaxies should be smaller than 80\% if
$\epsilon_{BH}=2.\,10^{-3}$ or $f<95\%$ if $\epsilon_{BH}=10^{-3}$. The 
predictions for $p=30$ agree with observations in all the cases.\par
For $\alpha=3/2$ (i.e. the gas profile approximates a King profile) and
for both cosmological models, we find 
$\overline y_{PG}$ of about $10^{-6}$ to a few $10^{-6}$, a 
prediction compatible with $\overline y_{FIRAS}$. This result 
remains valid for all values of $p$ between 10 and 30 including the highest
boundaries of $f$ and $\epsilon_{BH}$, respectively, 100\% and $2.\,10^{-3}$.
\subsection{Predicting the angular power spectrum}
\begin{figure}
\epsfxsize=\columnwidth
\hbox{\epsffile{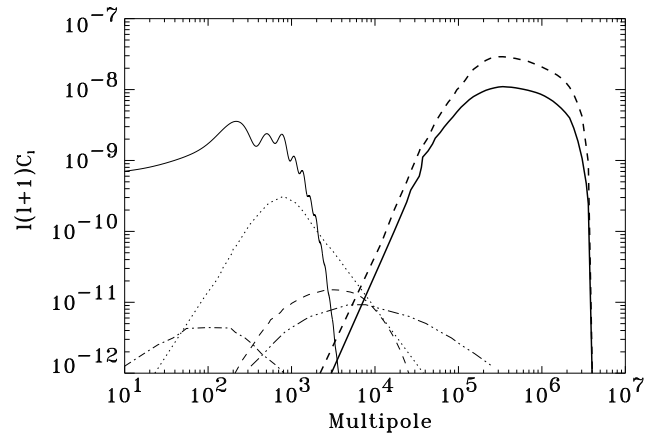}}
\caption{{\small\it Power spectrum of the temperature anisotropies. 
The solid thin line represents the CMB primary anisotropies. The solid thick
line represents the SZ kinetic effect of the proto-galaxies for $\Omega=1$. 
The thick dashed line is obtained for $\Omega=0.3$. The 
triple-dotted-dashed line represents the Vishniac-Ostriker effect. The dashed 
and dotted-dashed lines represent respectively the galaxy cluster contribution 
due to the kinetic SZ effect and the Rees-Sciama effect.}}
\label{fig:spec}
\end{figure}

We choose the set of parameters associated with the isothermal profile
which agrees with the COBE-FIRAS limit: $\alpha=1$, $p=30$, $f=95$\% and
$\epsilon_{BH}=2.\,10^{-3}$. Within this context, we predict the upper limit 
on the contribution to secondary 
temperature anisotropies induced by the SZ, thermal and kinetic effects, of the
proto-galactic gas. We express this contribution in 
terms of an angular power spectrum plotted in figure \ref{fig:spec} together 
with the main other well-known secondary anisotropies. \\
At very small scales ($l\sim$ a few
$10^5$) corresponding to galactic scales, the kinetic SZ contribution 
of the shock-heated gas (Fig. \ref{fig:spec}, thick solid line for
$\Omega_0=1$ and thick dashed line for $\Omega_0=0.3$) is very large. 
It is interesting to note the good agreement between our results and those 
obtained by \cite{peebles98a} 
for the scattering of the CMB photons by the cloudy proto-galactic plasma. 
 The power spectrum of the kinetic SZ anisotropies for the $\Omega_0=0.3$
model is significantly larger than the $\Omega_0=1$ model. This is mainly due
to the higher number of sources per unit comoving volume in open models. In 
all other flat cosmological
models involving a non-zero cosmological constant, the power spectrum will lie
between the two curves.  
The expected power spectrum due to the thermal SZ effect is not plotted in 
this figure. It is more than one order of magnitude smaller than the kinetic 
effect contribution. This is due to the efficiency of 
bremsstrahlung which lowers the temperature down to a few $10^5$ K. \par
We compare the contribution of the proto-galaxies due to their SZ kinetic 
effect to the major sources of secondary temperature anisotropies. In each 
case, we choose 
the most extreme cases for the comparison with our upper limit prediction. The 
power spectra displayed in figure \ref{fig:spec} are taken from the 
literature. The dotted line represents the upper limit of the contribution of 
the inhomogeneous reionisation as computed by \cite{aghanim96} 
for a quasar lifetime of $10^7$ yrs. The dot-dashed line
represents the Rees-Sciama effect \cite[]{rees68} 
taken from \cite{seljak96a} 
($\sigma_8=1$, $\Omega_0\,h=0.25$). The dashed line represents the galaxy
cluster contribution due to kinetic SZ effect from \cite{aghanim98}
with a cut-off mass of $10^{14}\,M_{\sun}$. The 
triple-dot-dashed line represents the Vishniac-Ostriker effect 
\cite[]{ostriker86,vishniac87} computed by \cite{hu96} 
with a total reionisation occurring at $z_i=10$. 
Finally, the solid thin line represents the power spectrum of the primary
CMB anisotropies in a standard CDM model computed using the CMBFAST code
\cite[]{seljak96}. 
The primary CMB anisotropies dominate at all scales
larger than the damping around 5 arcminutes. At 
intermediate scales, several effects take place among which the inhomogeneous
reionisation, the Ostriker-Vishniac and the SZ effect. In figure 
\ref{fig:spec}, we do not plot the power spectra of the thermal SZ effect of 
galaxy clusters. It is about one order of magnitude larger than the kinetic
SZ effect. At very small scales, 
the anisotropies are totally dominated by the proto-galactic contribution. 

\section{Conclusions}
Previous studies on galaxy formation have evaluated the global distortion of
a population of galaxies in the virialised regime. In these studies, the 
global Compton parameter was found to be very small, and smaller than the 
COBE-FIRAS value. In contrast, our model focuses on a regime in which  
proto-galaxies undergo a BH formation phase.
During this phase, the proto-galactic matter is shock-heated up to a few $10^8$
K and cools down to $10^5$ K.
CMB photons undergo inverse Compton scattering on the heated gas. In addition, 
galaxy
peculiar motions induce temperature anisotropies through the SZ kinetic effect.
We have estimated the global Compton parameter due to a population of 
proto-galaxies
and the expected power spectrum of the induced secondary anisotropies.
We find that there are four main parameters that control our model: the
fraction $f$ of BH-seeded proto-galaxies, the fraction $\epsilon_{BH}$ of the 
spheroid mass in the BH, the steepness of the density profile $\alpha$ and the 
gas core radius $p=R_{vir}/R_c$. The comparison between 
our predictions and the COBE-FIRAS observation constrains these parameters. 
Given the observed fraction of seeded galaxies, $f=95$\%, our results put 
rather strong 
constraints on the density profile and on $\epsilon_{BH}$. Indeed, our 
predictions agree with the observations whatever $p$ if the
density profile is an approximation to a King profile. On the contrary, if the
density profile is isothermal, then the core radius must be at least 30 times
smaller than the virial radius and the BH-to-spheroid mass ratio has to be
small, of the order of $10^{-3}$. The computations in the two extreme 
cosmological models show that the global Compton parameter due to 
proto-galaxies is not very sensitive to $\Omega_0$.

We compare the power spectra of the different contributions to the 
temperature anisotropies. Our results show that the SZ effect of the very 
early shock-heated proto-galaxies could constitute the major source of CMB 
distortions on very small scales (arcsecond and sub-arcsecond scales). 
The anisotropies 
are likely to be detected and measured by future long baseline 
interferometers such as ALMA. The shock heating is likely to contribute to the
re-heating
of the proto-galactic gas, which plays a role in galaxy formation theory. 
\cite{blanchard92} 
used preheating to modify the galaxy luminosity function, suppressing and 
finally delaying dwarf galaxy formation. We do not take into account 
this effect in our model, therefore, our results should be 
taken as an upper limit to the proto-galaxy contribution in terms of secondary 
anisotropies.

\begin{acknowledgements}
This work was initiated while C.B. was at the Center for Particle Astrophysics 
and while N.A. was at the 
Astronomy Department of the University of Berkeley (U.S.A.). N.A. acknowledges
partial funding from the Centre National d'\'Etudes
Spatiales. During this work, J.S. was partially supported by a Blaise
Pascal professorship. The authors thank O. Forni, G. Lagache and J.-L. Puget 
for discussions and careful reading of the manuscript. They also wish to thank
R. Juszkiewicz for his comments that helped to improve the final version of 
the paper.
\end{acknowledgements}

\end{document}